# Method for organizing the multiwavelength data of radio-loud active galactic nuclei

R. Schlickeiser[1] and C. D. Dermer[2]

[1] Max-Planck-Institut für Radioastronomie, Postfach 2024, D–53010 Bonn, Germany
[2] E. O. Hulburt Center for Space Research, Code 7653, Naval Research Laboratory, Washington, DC 20375-5352, USA

**Abstract.** The broadband emission observed from radio galaxies, BL Lac objects and quasars is thought to be produced by energetic §electrons confined in plasma blobs which are ejected from supermassive black holes at relativistic speeds. The lower energy radio-through-optical component is almost certainly nonthermal electron synchrotron radiation, while the higher-energy $\gamma$-ray component is probably due to Compton scattering of target photons by these same electrons. If the high-energy component is formed by jet electrons Compton-scattering radiation from outside the jet, such as the direct or rescattered accretion-disk photons, then the ratio of the power in the high-energy Compton component to the power in the low-energy synchrotron component stands in a simple relation depending on the observing angle to the jet and the plasma outflow speed. When combined with contemporaneous VLBI measurements of apparent transverse speed, we find that a simple diagram relates different classes of radio-loud active galactic nuclei (AGNs) and makes definite predictions for multiwavelength observations of these sources. If an equipartition parameter remains constant between episodes of plasma ejection, then multiple observations of a single source can in principle determine the Hubble constant.

## 1. Introduction

The telescopes onboard the *Compton Gamma Ray Observatory* have made possible the detection of more than 65 active galactic nuclei (AGNs) at photon energies $E \geq 0.05$ MeV (Fichtel et al. 1994; Kurfess 1994; Montigny et al. 1995; Dermer & Gehrels 1995). Fifty-two of these sources are radio-loud, and except for five radio galaxies and two BL Lac objects so far detected only at $E < 1$ MeV, the remainder are associated with AGNs which are classified in other wavelength ranges as blazars. Because of the rapid variability (timescales of days) of the $> 100$ MeV emission from many $\gamma$-ray blazars (Michelson et al. 1994), their large fluxes and inferred high compactnesses, and their association with radio-loud sources, it is generally agreed (Dermer & Schlickeiser 1992; Blandford 1993) that the observed emission at radio, mm, optical and $\gamma$-ray frequencies originates in strongly beamed sources, in accordance with the relativistic beaming hypothesis that has served well as the baseline model for the central engines of AGNs (Blandford & Rees 1978; Blandford 1990).

Apparent superluminal (SL) motion is also a common feature of $\gamma$-ray blazars, and $\approx 25\%$ of AGNs detected with the EGRET telescope on *Compton* at $E > 100$ MeV have been reported as SL sources. The connection between blazar $\gamma$-ray production and apparent SL motion was first pointed out by Dermer, Schlickeiser, & Mastichiadis (1992), and the recent VLBI observations of PKS 0528+134 by Pohl et al. (1995), showing an apparent SL speed $v \equiv \beta c = 4.4(\pm 1.7) c h^{-1}$, where the Hubble constant $H_0 = 100 h$ km s$^{-1}$ Mpc$^{-1}$, supports a connection between blazar $\gamma$-ray emission and apparent SL motion.

However, 3 of the 5 radio galaxies detected with the OSSE instrument on *Compton* at photon energies between 0.05 and 1 MeV, namely 3C 111, 3C 120 and 3C 390.3, also show strong evidence for SL motion. Cen A and 3C 84 (NGC 1275), the other two radio galaxies detected with OSSE, have measured subluminal components. Although Cen A has been detected with COMPTEL in the 0.7-1.5 MeV range, none of the 5 radio galaxies has been positively detected at $E > 100$ MeV with EGRET, so that measurement of SL motion does not necessarily imply that a given AGN strongly emits $\gamma$-rays in the EGRET range.

It is the purpose of this Letter to demonstrate that a suitable combination of multiwavelength observations and VLBI measurements of the speed of newly emerging jet components (hereafter referred to as "blobs") from the central black hole provides a unique tool to determine key physical parameters of blazar jet models and, if more than two new jet components are detected, the cosmological Hubble constant.

## 2. Relation between Compton-synchrotron parameter and jet speed

According to the unification scenario for radio-emitting AGNs (e.g. Woltjer 1990; Antonucci 1993; Urry & Padovani 1995), blazars are radio galaxies viewed nearly along the axis of a jet of relativistically outflowing plasma. The fact that radio galaxies are not EGRET sources suggests that the $\gamma$ radiation is more beamed than the synchrotron emission. According to models where relativistic electrons are the primary radiating particles and photons external to the jet are Compton scattered by these jet electrons (Dermer & Schlickeiser 1993; Sikora, Begelman & Rees 1994; Blandford & Levinson 1995), the ratio of the $\nu F_\nu$ spectral power in the synchrotron and Compton components is given by

$$\rho \simeq k D^{1+\alpha} \qquad (1)$$

(Dermer, Sturner & Schlickeiser 1995, Eq. (27); Dermer 1995), where $k = u_{iso}/u_B$ denotes the ratio of the energy densities in the external target radiation field and the blob's magnetic field, while $\alpha$ is the measured energy spectral index of the Compton scattered photon power law flux. The Doppler factor

$$D = [\Gamma(1 - B\mu)]^{-1}, \qquad (2)$$

where $\Gamma = (1 - B^2)^{-1/2}$ is the Lorentz factor of the outflowing plasma blob, $Bc$ denotes the blob's bulk velocity, and $\theta = \cos^{-1}\mu$ denotes the observer's viewing angle with respect to the jet axis. We can simply invert equation (1) to obtain $D$ in terms of the measured values of $\rho$ and $\alpha$, giving

$$D = \left(\frac{\rho}{k}\right)^{1/(1+\alpha)} \equiv \chi. \qquad (3)$$

The relation between the measured apparent superluminal transverse speed $\beta c$, the blob's speed $Bc$, and the observing angle $\theta$ is given by

$$\beta = \frac{B\sqrt{1-\mu^2}}{1 - B\mu} \qquad (4)$$

(see Rees 1966; Rybicki & Lightman 1979). Eqs. (3) and (4) can be combined into the expression

$$\theta = \arctan\left[\frac{2\beta}{\chi^2 + \beta^2 - 1}\right]. \qquad (5)$$

In Fig. 1 we show the relationship between $\beta$ and the Compton/synchrotron parameter $\chi$ for different values of $\theta$ and $\Gamma$ (see also Unwin et al. 1983). By following lines of constant $\Gamma$, we see that as $D$ or $\chi$ increases, $\beta$ rises from 0 at large values of $\theta$ to its maximum value ($= B\Gamma$) at the observing angle $\theta_{SL} = \cos^{-1} B$, returning to small values at small observing angles. For a value $\Gamma = 10$, which may be typical of quasars, moderate values of $\beta$ between 0 and 5 occur when viewing at large angles $\theta \gtrsim 15°$ to the jet axis. This is when $D$ or $\chi$ is small, so we expect a small ratio of $\gamma$-ray power to synchrotron power. At shallow angles $\theta \simeq 3°\text{-}10°$, SL speeds $\beta \approx B\Gamma$ will be observed, corresponding to moderate or large values of $\chi$. For the rare cases where we are viewing a few degrees or less to the jet axis, large values of $\chi$ should accompany small values of $\beta$.

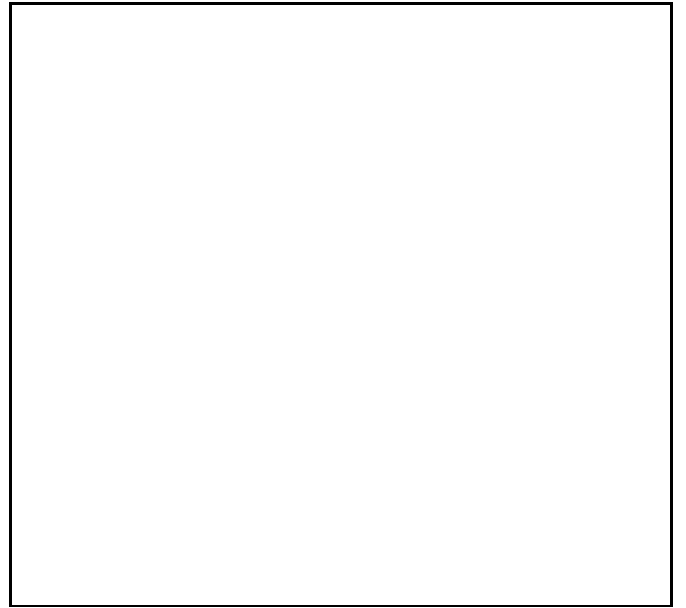

**Fig. 1.** Relation between apparent transverse speed $\beta$ and the Compton/synchrotron parameter $\chi$ for different values of the observing angle $\theta$ with respect to the jet direction (solid curves) and the bulk Lorentz factor $\Gamma$ (dashed curves). The innermost solid curve is calculated for $\theta = 45°$ and the innermost dashed curve for $\Gamma = 2$.

Supermassive black holes have considerable gyroscopic stability and would therefore eject plasma along fairly well-defined axes. We can thus chart the expected relation between $\beta$ and $\chi$ for a given source. This is given by curves of constant $\theta$ in Fig. 1, where we see that a single source ejecting plasma along a fixed axis but with variable values of $\Gamma$ would display values of $\beta$ and $\chi$ that follow a prescribed trajectory for a fixed value of the equipartition value $k$. The constancy of $k$ is a crucial, but testable, assumption in our model. We note, however, that Scott & Readhead (1977) find no evidence for departures from equipartition between particle and magnetic field energy densities.

A definite test of these kinematic relations based upon an external Compton-scattering model for the $\gamma$-rays will require correlated VLBI, mm-through-optical monitoring of the synchrotron component, and γ-ray monitoring of the Compton component. At present, there are only a few correlated multiwavelength campaigns which monitor both the synchrotron emission above strongly self-absorbed frequencies ($\gg 100$ GHz) and the Compton component at γ-ray energies, including 3C 273 (Lichti et al. 1995), 3C 279 (Maraschi et al. 1994) PKS 0528+134 (Pohl et al. 1995), and Mrk 421 (Macomb et al. 1995). For PKS 0528+134 (Krichbaum et al. 1995) and 3C 279 (Wehrle et al. 1994), correlated VLBI studies have been performed, but final results are not yet available. To test the implications of this method, we therefore consider the results for non-contemporaneous measurements of many sources.

## 3. Analysis of non-contemporaneous multiwavelength data

In Fig. 2, we plot non-contemporaneous data for $\chi$ and $\beta$. For the γ-ray data, we use the all-sky average $> 100$ MeV flux from the EGRET Phase 1 catalog (Fichtel et al. 1994), to which we include the average γ-ray flux during the recent outburst of 1156+295 (Webb et al. 1995). Only upper limits of the $> 100$ MeV γ-ray flux have been reported for the radio galaxies 3C 84, 3C 120 and Cen A (Fichtel et al. 1994). For the galaxies 3C 111 and 3C 390.3, we use a conservative upper limit of $2 \times 10^{-7}$ ph cm$^{-2}$ s$^{-1}$. Where available, we use the mm data at 375 GHz from Bloom et al. (1994). In other cases, we extrapolate from lower frequencies (Reich et al. 1993; Owen & Puschell 1982; Owen, Spangler, & Cotton 1980). For 1156+295 and 1633+382, upper limits to the 375 GHz flux result in lower limits for the Compton/synchrotron parameter. For the apparent transverse speeds we use the reported values in the literature (Impey 1987; Vermeulen & Cohen 1994; Dermer & Gehrels 1995). Multiple measurements of $\beta$ for a single source are treated as individual data points.

Fig. 2 reveals some very interesting features. The radio galaxies group in the lower left corner of the diagram, whereas the γ-ray quasars have, in general, larger values of $\beta$ and $\chi$. Comparing with Fig. 1, this behavior can be understood if we are viewing at large angles with respect to the jet axis for radio galaxies, and at small angles for blazars. This is consistent with the unification scenario for radio-loud AGNs, provided that the γ rays are are produced by an external Compton scattering process. A more stringent test of this hypothesis will require contemporaneous mm-through-optical and EGRET γ-ray observations, combined with high resolution VLBI measurements of $\beta$. We recognize that the measurements of $\beta$ take place months to years after the synchrotron/Compton flare that probably accompanies the emergence of a new jet component (Wehrle et al. 1994), so that we must assume that the blob is not accelerated following the outburst. This appears to be true for most plasma outflows monitored during different VLBI campaigns of a single source.

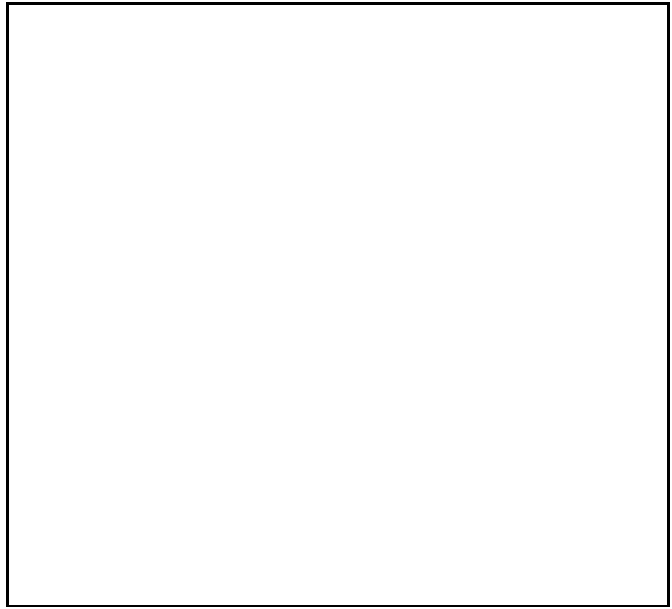

**Fig. 2.** Observed relation between apparent transverse speed $\beta$ and the Compton/synchrotron parameter $\chi$ for γ-ray emitting quasars (crosses), radio galaxies (circles), and the BL Lac objects Mrk 421 and 0235+164 (diamonds) from non-contemporaneous data, assuming a constant value of the equipartition parameter $k = 1$. A source with more than one reported value of $\beta$ appears as several data points.

Two BL Lac objects also appear in Fig. 2. Mrk 421 has an inferred moderate to large value of $\chi$, but small measured values of $\beta$. By contrast, the source 0235+164 has a large measured value of $\beta$, but such large values of $\beta$ from this source needs to be confirmed by future measurements (the measured $\beta$ does not appear in the list by Vermeulen & Cohen 1994). It is interesting to note that the location of both BL Lac objects in the diagram are consistent with small viewing angles ($\theta \lesssim 5°$; see Fig. 1). This result is in possible conflict with the interpretation of Padovani & Urry (1991, 1992), who argue that FR 1 galaxies are the parent populations of BL Lac objects, and that the values of $\Gamma$ for BL Lac objects are $\lesssim 7$. But it is important to note that the peak of the synchrotron component of Mrk 421 is at considerably higher frequencies, so that $\chi$ must be smaller. Future campaigns to monitor BL Lac objects can test whether the data is more consistent with small jet angles or with small values of $\Gamma$.

## 4. Determination of source parameters and the Hubble constant

Of particular interest are γ-ray blazars that undergo several outbursts. Lately it has been established for the blazars 0534+134 (Pohl et al. 1995; Krichbaum et al. 1995), 3C 279 (Wehrle et al. 1995), 3C 454.3 (Krichbaum et al. 1995) and 1633+382 (Barthel et al. 1995) that en-hanced levels of activity in the optical and $\gamma$-ray bands are associated with the emergence of new jet components. Assuming that the observing angle $\theta$, the equipartition parameter $k$ and the energy spectral index $\alpha$ entering Eq. (3) remain constant at different blob ejections, then observations of two different blobs can uniquely determine the values of $\theta$ and the equipartition parameter $k$. For blazars with two emerging outbursts and measured values of $(\beta_1, \rho_1)$ and $(\beta_2, \rho_2)$, we infer that the equipartition parameter has the value

$$k = \left[\frac{\beta_1 \rho_2^{2/(1+\alpha)} - \beta_2 \rho_1^{2/(1+\alpha)}}{(\beta_1 - \beta_2)(1 + \beta_1 \beta_2)}\right]^{(1+\alpha)/2}, \quad (6)$$

and that the observing angle is given by

$$\theta = \arctan\left[\frac{2(\beta_2 \rho_1^{2/(1+\alpha)} - \beta_1 \rho_2^{2/(1+\alpha)})}{(\beta_2^2 - 1)\rho_1^{2/(1+\alpha)} - (\beta_1^2 - 1)\rho_2^{2/(1+\alpha)}}\right]. \quad (7)$$

The determination of the apparent speed $\beta$ depends on the Hubble constant $H_0 = 100h$ km s$^{-1}$ Mpc$^{-1}$ (e.g. Pearson & Zensus 1987) through the relation

$$\beta = \frac{95}{h}[1 - \frac{1}{\sqrt{1+z}}](\frac{T}{\text{mas/yr}}), \quad (8)$$

given the measured proper motion $T$ for a $\Omega = 1$, $\Lambda = 0$ universe. Under the same assumptions given above, the observation of a third ejection can determine the value of $h$.

## 5. Conclusions

If the $\gamma$ radiation from radio-loud active galactic nuclei is generated by relativistic electrons in the jet which Compton scatter radiation from outside the jet, such as the direct or rescattered accretion-disk photons, then the ratio of the power in the high-energy Compton component to the power in the low-energy synchrotron component is simply related by Eq. (1), which depends on the observing angle to the jet and the plasma outflow speed. This implies that the $\gamma$ rays are emitted into a narrower cone than the synchrotron radiation (Dermer 1995). We have demonstrated that the combination of this simple relation with contemporaneous VLBI measurements of apparent transverse speed leads to a simple diagram that relates different classes of radio-loud AGNs and makes definite predictions for multiwavelength observations of these sources. For observations of a single source at different episodes of plasma ejection, we can test our assumption that the equipartition parameter remains constant, determine the angle between the observer and the jet axis and, in principle, measure the Hubble constant.

*Acknowledgements.* RS gratefully acknowledges partial support by the DARA (50 OR 9406 3) of his Compton observatory guest investigator programs. CD acknowledges the kind hospitality at the MPI für Radioastronomie where this work was completed.

**Fig. 1** Relation between apparent transverse speed $\beta$ and the Compton/synchrotron parameter $\chi$ for different values of the observing angle $\theta$ with respect to the jet direction (solid curves) and the bulk Lorentz factor $\Gamma$ (dashed curves). The innermost solid curve is calculated for $\theta = 45°$ and the innermost dashed curve for $\Gamma = 2$.

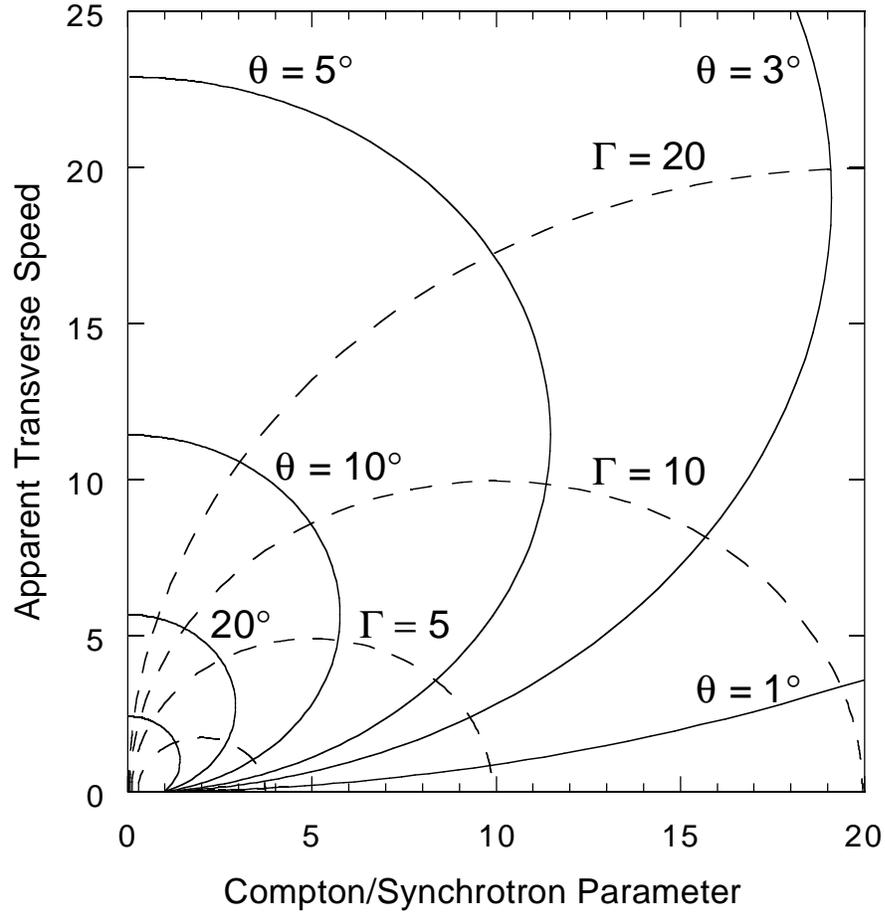

**Fig. 2** Observed relation between apparent transverse speed $\beta$ and the Compton/synchrotron parameter $\chi$ for $\gamma$-ray emitting quasars (crosses), radio galaxies (circles), and the BL Lac objects Mrk 421 and 0235+164 (diamonds) from non-contemporaneous data, assuming a constant value of the equipartition parameter $k = 1$. A source with more than one reported value of $\beta$ appears as several data points.

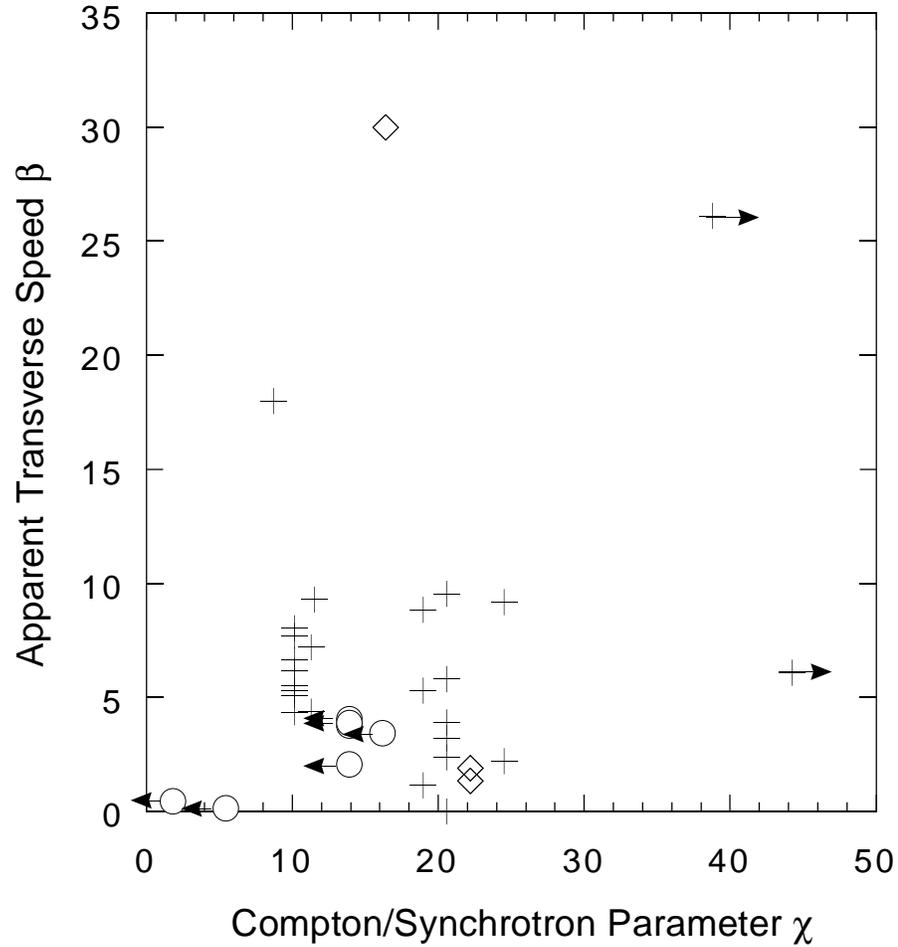